\documentclass{elsart}
\usepackage{graphicx}
\usepackage{amssymb}

\begin{document}

\begin{frontmatter}



\title{Measurements of anisotropic scintillation efficiency for carbon
 recoils in a stilbene crystal for dark matter detection}


\author[phys]{Hiroyuki Sekiya\corauthref{cor}},
\corauth[cor]{Corresponding author.}
\ead{sekiya@icepp.s.u-tokyo.ac.jp}
\author[phys]{Makoto Minowa},
\author[phys]{Yuki Shimizu},
\author[icepp]{Yoshizumi Inoue},
\author[phys]{Wataru Suganuma}

\address[phys]{Department of Physics, School of Science, University of Tokyo,
7-3-1, Hongo, Bunkyo-ku, Tokyo 113-0033, Japan}

\address[icepp]{International Center for Elementary Particle Physics (ICEPP),
 University of Tokyo, 7-3-1, Hongo, Bunkyo-ku, Tokyo 113-0033, Japan}

\begin{abstract}
% Text of abstract
It is known that scintillation efficiency of organic single crystals
 depends on the direction of nuclear recoils relative to
 crystallographic axes. This property could be applied to the
 directional WIMP dark matter detector.

The scintillation efficiency of carbon recoils in a stilbene crystal
 was measured for recoil energies of 30 keV to 1 MeV using neutrons from 
$^7$Li(p,n)$^7$Be and  $^{252}$Cf.
Anisotropic response was confirmed in low energy regions. The  variation
 of the scintillation efficiency was about 7 \%, that could detect the possible
dark matter signal.

\end{abstract}

\end{frontmatter}
 
% main text
\section{Introduction}
\label{intro}
It is considered that the galactic halo is 
composed of weakly interacting massive particles (WIMPs) as dark matter.
These particles could be directly detected by measuring 
the nuclear recoils produced by their elastic scattering off nuclei
in detectors \cite{jungman}.
However, nuclear recoils produced by background neutrons are
indistinguishable from those by WIMPs.
Therefore, one should look for statistical signature of the WIMPs.
Realistic distinctive WIMP signals arise from the
earth's motion  in the galactic halo.
Annual modulation of event rate is one of the possible WIMP signals
caused by earth's revolution around the Sun ($\sim$30 km/s). 
Actually, this signature was argued by the DAMA experiment \cite{DAMA}.
  
Nevertheless, the most convincing signature of the WIMPs appears in the
directions of nuclear recoils. It is provided by the earth's velocity
through the galactic halo ($\sim$230 km/s). 
Assuming the  WIMP halo is an isothermal sphere,
 the strong WIMP wind is blowing on the earth and the distribution
of the nuclear recoil direction shows a large asymmetry.
Hence, detectors sensitive to the direction
of the recoil nucleus would have a great potential to identify WIMPs, and
further, to provide information on the galactic halo \cite{halo2}. 

It is known that scintillation efficiency of organic crystals
to heavy charged particles depends on the direction of the particles 
with respect to the crystallographic axes \cite{birks,heckmann}.
This property makes it possible to propose a WIMP detector sensitive to
the recoil direction of the nucleus \cite{belli,uk}.
The directional scintillation efficiency to recoil protons and carbons 
produced by neutrons in MeV regions was reported \cite{knoll}, 
but the recoil energy given by WIMPs is much lower. 

We had measured the proton recoils in a stilbene crystal for recoil
energies of 300 keV to 3 MeV and shown that the efficiency depends
 on the direction of the recoil proton \cite{pikachu}. 
However, carbon recoils would be more effective than proton recoils in 
 detecting WIMPs with the interesting mass region of above 10 GeV.
In this paper, we report on the anisotropic scintillation efficiency of
carbon recoils in a stilbene crystal with neutrons especially in low energy regions.
We also estimate the sensitivity to the WIMP wind of the stilbene 
crystals.

\section{Experimental Setup}
\label{Experiments}
\begin{figure}[p]
\begin{center}
\includegraphics{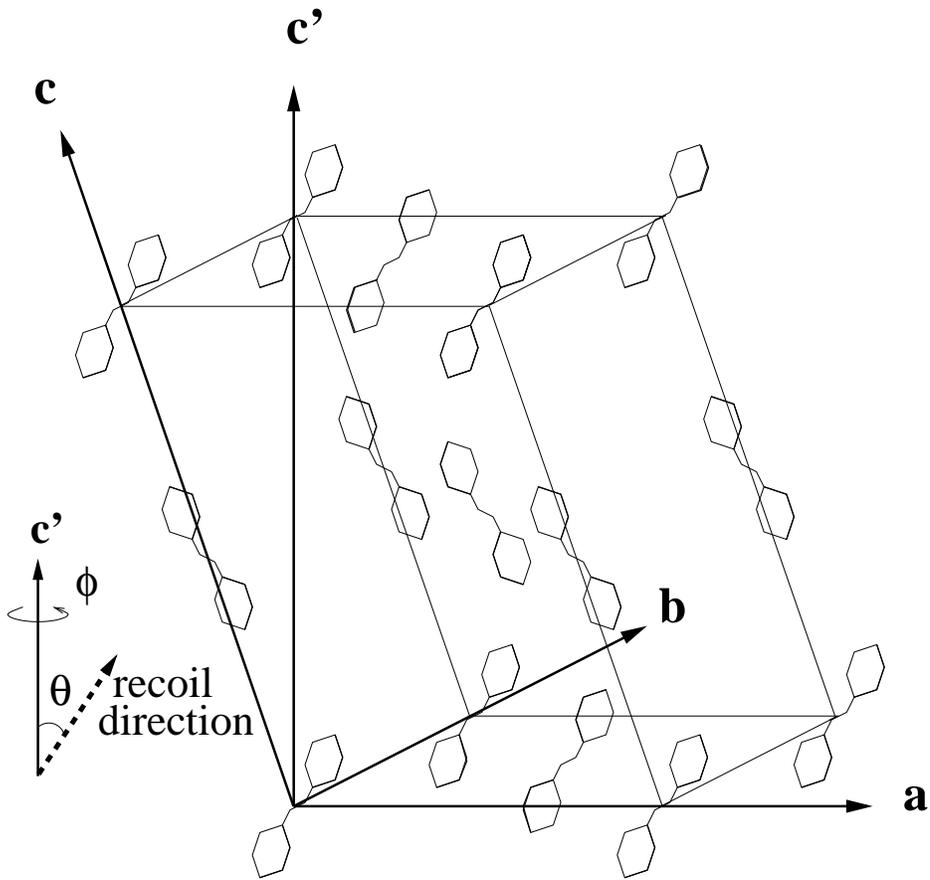}
\caption{Schematic drawing of the stilbene crystal lattice and the definition of
 the recoil angles. $\theta$ is the recoil angle with respect to the
 $c'$ axis and $\phi$ is the angle around the $c'$ axis.}
\label{fig:crystil}
\end{center}
\end{figure}

The crystal lattice of stilbenes is shown in Fig. \ref{fig:crystil}. 
Stilbene crystals form monoclinic systems and the
crystallographic axes are called $a$, $b$, and  $c$. 
The axis perpendicular to $a$-$b$ plane is called $c'$.
The direction of $c'$ axis is easily discerned because stilbene crystals
are cleaved along $a$-$b$ plane, and according to \cite{heckmann}, the scintillation
efficiency of stilbene crystals depends on the recoil angle 
with respect to $c'$. Therefore, we measured the recoil angle dependence
by changing the angle of the recoil direction with respect to $c'$ axis
($\theta$) and the angle around the $c'$ axis ($\phi$, $\phi=0^{\circ}$ was
determined arbitrarily).

\begin{figure}[p]
\begin{center}
\includegraphics{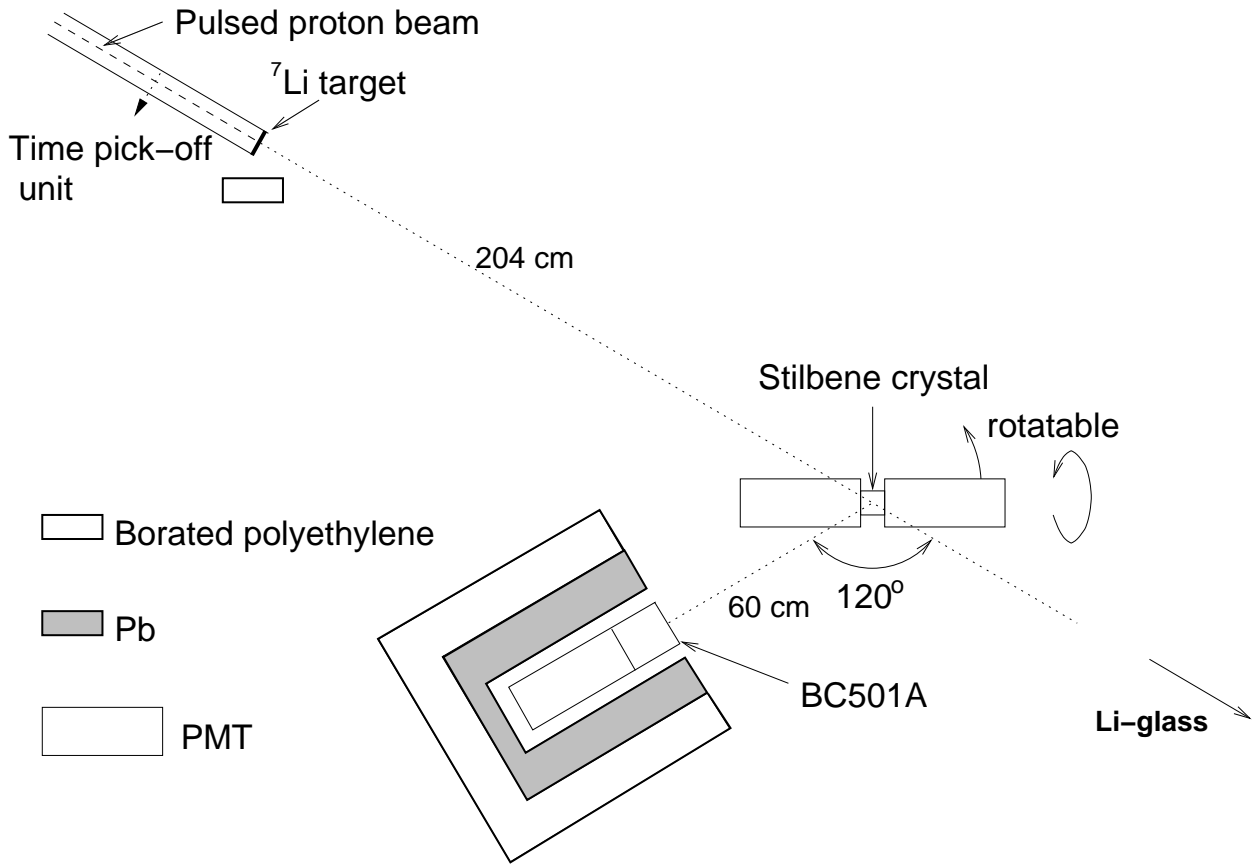}
\caption{Setup of the $^7$Li(p,n)$^7$Be source run. Because the crystal
 shape is cubical and $c'$ is perpendicular to cleavage planes where
PMTs are attached, the geometrical configuration of the crystal relative 
to the neutron beam is same for $\theta=90^{\circ}$ and $\theta=0^{\circ}$.  }
\label{fig:setup}
\end{center}
\end{figure}

In order to obtain high statistics and various incident neutron energy,
two neutron sources, $^7$Li(p,n)$^7$Be and $^{252}$Cf  were employed.
The $^7$Li(p,n)$^7$Be source run was performed at 3.2 MV 
Pelletron accelerator of the Research Laboratory for Nuclear
 Reactors at Tokyo Institute of Technology. 
Pulsed proton beam interacted with a thin lithium target, and
pulsed neutrons were produced by the $^7$Li(p,n)$^7$Be reaction. 
The repetition rate of the pulsed proton beam was set at 2 MHz and
the pulse width was 1.5ns. 
Changing proton energy three times, neutrons with energies of 200 to 650
keV were obtained.  The beam and detector geometry is illustrated in
Fig. \ref{fig:setup}.
The temperature of the room was kept at 26$^{\circ}$C during the measurement.

The dimension of the target stilbene that we measured
was 2 cm $\times$ 2 cm $\times$ 2 cm. Two opposite faces were
cleaved ($a$-$b$ plane) and other four sides (arbitrary plane) 
were polished. 
The polished sides were covered with
GORE-TEX$^{\mbox{\scriptsize{\textcircled{\tiny R}}}}$, and 51 mm $\phi$ PMT
(Hamamatsu H6411) was attached to each cleavage plane. Self-coincidence
between two PMTs was required to reduce dark current events.
Scattered neutron was detected by 51 mm $\phi$ $\times$ 51 mm liquid
scintillator encapsulated in an aluminum cell (Saint-Gobain BC501A-MAB1)
with PMT (Hamamatsu H6411). 
The scattering angle was fixed at 120$^{\circ}$ to avoid proton
recoil events. 

The incident and scattered neutron energies were measured by
time-of-flight (TOF) method and recorded by a TDC (Hohshin C021).
The coincident stilbene output was used to define the common ``START''.
The outputs of the BC501A detector  and the delayed pulsed proton signal from  the
``Time Pick-off Unit'' of the accelerator were used as the ``STOP'' signals.
The PMT outputs were recorded by charge ADCs (Hohshin C009H) with two
different gates ---
 20 ns and 500 ns from the rise of the waveforms of the PMT outputs.
Both stilbene and BC501A provide pulse shape discrimination (PSD)
capabilities for n-$\gamma$ separation arising from the difference of 
the slow scintillation component fraction \cite{birks}.
Consequently, the ratio of charge integrated by two different gates
gives n-$\gamma$ information.

The setup of the $^{252}$Cf source run was almost the same as 
the $^7$Li(p,n)$^7$Be source run, except the measurements of incident
neutron energy. The target stilbene crystal was placed 60 cm away from
the 3.5 MBq  $^{252}$Cf and the BC501A detector was placed 60 cm away
from the target at 120$^{\circ}$ of the scattering angle.
To determine the timing of the nuclear fission of 
$^{252}$Cf, prompt gamma rays produced by the fission  were detected
with the 2 cm thick plastic scintillator located  by the neutron source. 
PMT (Hamamatsu R1250) outputs were delayed and used as the ``STOP''
signal of the TDC.

\section{Measurement results}
\label{results}
\begin{figure}[p]
\begin{center}
\includegraphics[width=12cm]{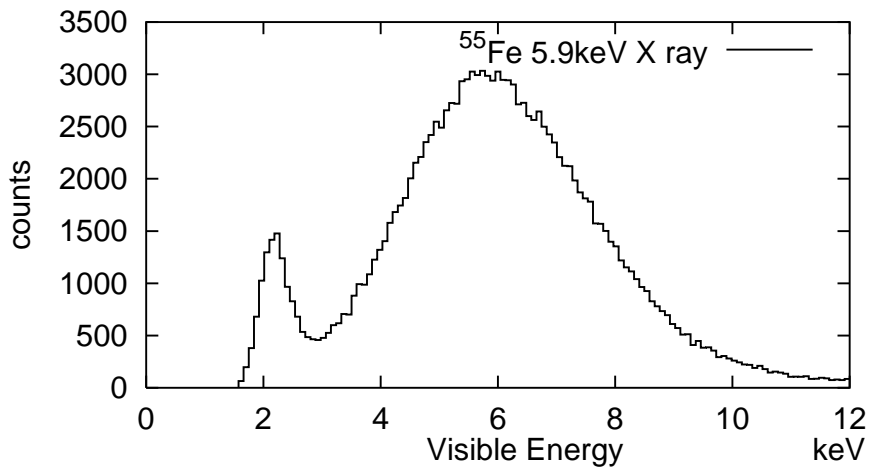}
\caption{The energy spectrum for 5.9 keV X rays  of $^{55}$Fe obtained
 from the stilbene crystal.}
\label{fig:fe}
\end{center}
\end{figure}

The detected electron equivalent energy (visible energy) for carbon recoils in
the stilbene crystal was calibrated with 5.9 keV, 6.4 keV, 14.4 keV,
22.2 keV, and 32.2 keV X/gamma rays from $^{55}$Fe,$^{57}$Co,$^{109}$Cd,
and $^{137}$Cs. The efficiency close to the threshold was checked 
by the 5.9 keV X rays. We see from Fig. \ref{fig:fe} that
the efficiency was not lost above 3keV.

Although about 40$\%$ of neutrons interacts with the stilbene crystal more
than twice in the measured energy region\footnote{Based on the GEANT3
simulation}, the single recoil events of
neutrons scattered at $120^{\circ}$ by carbons were selected using the
TOFs of the incident and scattered neutrons.  

However, in the $^{7}$Li(p,n)$^{7}$Be source run, there still remained
background events. These were gamma ray events from $^{7}$Li(p,$\gamma$)$^{8}$Be
(minority) and nuclear recoils in the stilbene crystal accidentally coincident with 
the events in the BC501A detector (majority).
The gamma ray events were rejected by means of PSD of the both stilbene and
BC501A detectors. The events around 0 ns of the TOF spectra of both incident 
neutrons and scattered neutrons are also rejected as the gamma ray events. 
On the other hand, to estimate the backgrounds of the accidental nuclear recoils, 
the measurements without requiring the coincidence of the BC501A
detector were performed before and after each of the $\theta$/$\phi$ run.
Those background spectra were stable throughout the experiment. 
The stability of the beam intensity was also confirmed by the Li glass
scintillator in the beam line.  
 
\begin{figure}[p]
\begin{center}
\includegraphics[width=12cm]{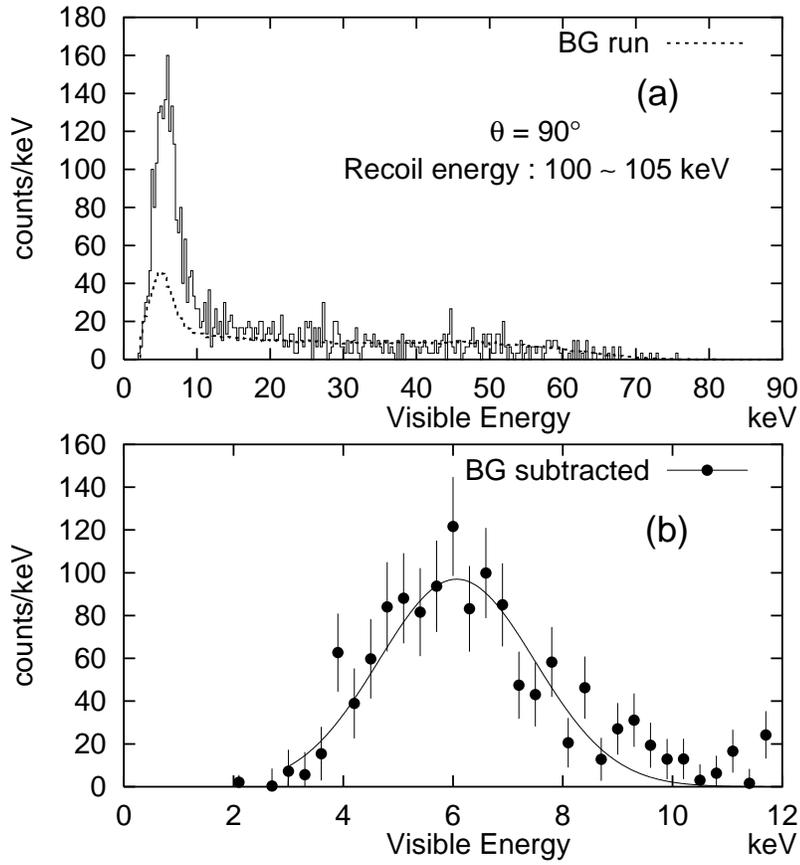}
\caption{(a) The visible energy spectra with $\theta=90^{\circ}$ for
recoil energies of 100 to 105 keV (solid) and the normalized
 background (dashed). (b) The visible energy spectrum after background
subtraction and the fitted Gaussian.}
\label{fig:analysis}
\end{center}
\end{figure}

The visible energy spectrum with $\theta=90^{\circ}$ for recoil
energies of 100 to 105 keV is shown in Fig. \ref{fig:analysis} together with
the background spectrum after gamma ray events rejection.
As indicated, background spectrum was normalized and subtracted.
The normalization factor was the ratio of the number of events
of the higher energy regions.
The shape of the background spectrum below 10 keV reflects the fact
that the scintillation efficiency of proton recoils increases at
low energies \cite{hong} (Not the loss in efficiency near the threshold).
The visible energy of that recoil energies was derived by 
fitting Gaussian to the spectrum above 3keV.

The analysis procedure of the $^{252}$Cf source run was detailed 
in \cite{pikachu}. After the events were binned with recoil energy,
the visible energy was determined by taking the average in each bin 
of the recoil energy because of the small statistics.

\begin{figure}[p]
\begin{center}
\includegraphics{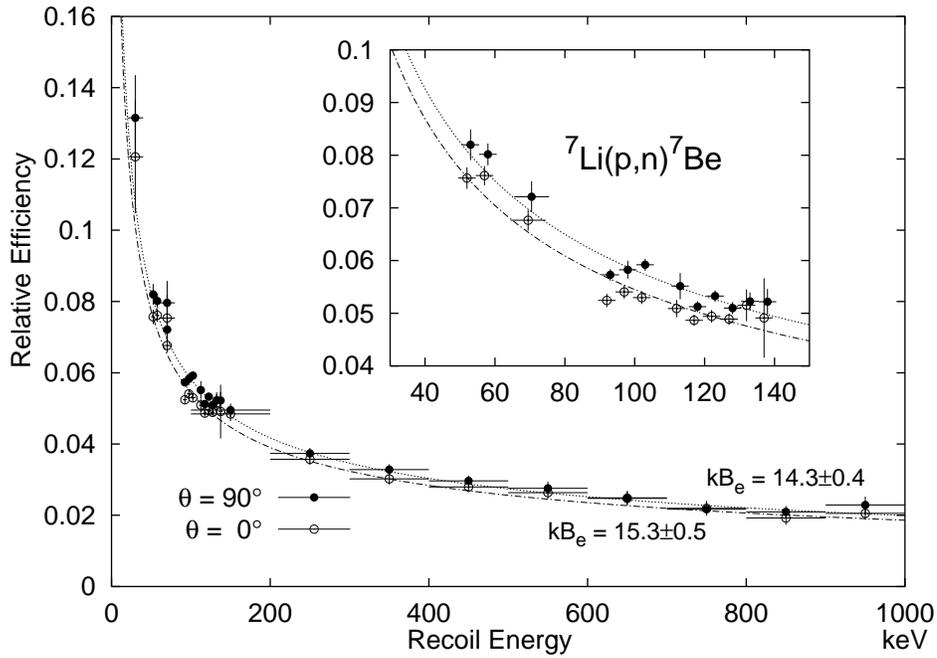}
\caption{The measured scintillation efficiency relative to the
 efficiency for electrons  with $\theta=0^{\circ}$ and
 $\theta=90^{\circ}$ recoils ($\phi=0^{\circ}$).  The lines show best
 fits to the data using
 equation (\ref{eq:modbirks}). $S_e/S_n$ is $0.547\pm0.030$ for the both
 cases. The inset is the results of the
 $^7$Li(p,n)$^7$Be source run. Horizontal error bars represent selected
 recoil energy region for calculating the efficiency. }
\label{fig:result1}
\end{center}
\end{figure}

The derived scintillation efficiency relative to the efficiency for electrons
with  $\theta=90^{\circ}$ and $\theta=0^{\circ}$ are shown 
in Fig. \ref{fig:result1}. 
The result is consistent with the measurements for proton
recoils \cite{pikachu} and for charged particles 
in high energy regions \cite{heckmann}, that is,
the scintillation efficiency of stilbene crystal for carbon recoil is
maximal in the direction perpendicular to the $c'$ axis
($\theta=90^{\circ}$) and minimal in the direction parallel to the $c'$
axis ($\theta=0^{\circ}$).
On the other hand, $\phi$-dependence of the scintillation efficiency 
was not observed in this measurement.
That is also consistent with \cite{heckmann}.

In order to estimate the response to the WIMPs, 
it is necessary to express the scintillation efficiency as a function of
the recoil energy.
The relation between the light yield ($dL/dx$) and the energy
loss ($dE/dx$) of scintilltors had been explained by Birks's empirical
model \cite{birks}.
However, this model does not describe the rise of the efficiency
at low energies of our results.
Recently, such enhancement at low energies for organic liquid
scintillator was reported \cite{hong}, and the light yield is introduced as
\newcommand{\fx}{\left(\displaystyle{\frac{dE}{dx}}\right)_e}
\newcommand{\gx}{\left(\displaystyle{\frac{dE}{dx}}\right)_n}  
\begin{equation}
\frac{dL}{dx}=\frac{S_e\fx+S_n\gx}{1+kB_e\fx+kB_n\gx}
\label{eq:modbirks}
\end{equation}
with $kB_e/kB_n=3250$.
This relation is based on the assumption that electronic energy loss and
nuclear energy loss contribute differently to the quenching process and
scintillating process.
The ($dE/dx$) terms are given by SRIM2003 package \cite{srim} as a
function of recoil energy. 

While $S_e/S_n$, $kB_e$, or both of them could be function of the
angle $\theta$ for this experiment,
 it is suggested that the response anisotropy is
due to the difference in the value of the ionization quenching
parameter $kB$ \cite{birks}.
Therefore, we fitted equation (\ref{eq:modbirks}) to both of the 
measured $\theta=90^{\circ}$ and $\theta=0^{\circ}$ data at a time
 with the three parameters; $kB_e(\theta=90^{\circ})$,
 $kB_e(\theta=0^{\circ})$, and $S_e/S_n$. 
The lines in Fig. \ref{fig:result1} are the best fits to the data with
$kB_e(\theta=90^{\circ})=14.3\pm0.4$ mg/cm$^2$/MeV,
$kB_e(\theta=0^{\circ})=15.3\pm0.5$ mg/cm$^2$/MeV and
$S_e/S_n=0.547\pm0.030$ ($\chi^2=68.7$, ${\rm d.o.f.}=43$).
Although statistics is not very high, anisotropy of 7$\%$ is seen over the
measured energy region.

\section{Discussion}
The WIMP signal by the stilbene scintillator can be obtained by
comparing the visible energy spectra measured for different orientation 
with respect to earth's motion in the galactic halo.
The variation amplitude (difference of the spectra) should be 
independent of any terrestrial backgrounds.
Assuming spherical isothermal halo model, we estimate the expected
variation of the visible energy spectrum.
The differential angular rate of WIMPs in detectors with respect to
laboratory recoil angle $\gamma$ and recoil energy $E_R$ with WIMP
mass $M_{\chi}$ and target nucleus mass $M_n$ is given 
by \cite{spergel}:
\begin{equation}
\frac{d^2R}{dE_R d\cos\gamma}=
\frac{\sigma\rho_0 (M_{\chi}+M_n)^2}{2\pi^{\frac{1}{2}}M_{\chi}^3M_{n}v_0}
\exp\left[-\frac{\left(v_E\cos\gamma-v_{\rm min}\right)^2}{v_0^2}\right],
\label{eq:dist}
\end{equation}
where $\rho_0$ is the local WIMP density, $v_0$ is the velocity
dispersion of the isothermal halo, and
$v_E$ is the component of the earth's velocity parallel to the galactic rotation;
$v_{\rm min}^2=\left(M_{\chi}+M_{n}\right)^2E_R/2M_{\chi}^2M_n$ is the minimum WIMP
velocity that can produce $E_R$ and $\sigma$ is the WIMP-nucleus cross section. 

\begin{figure}[p]
\begin{center}
\includegraphics{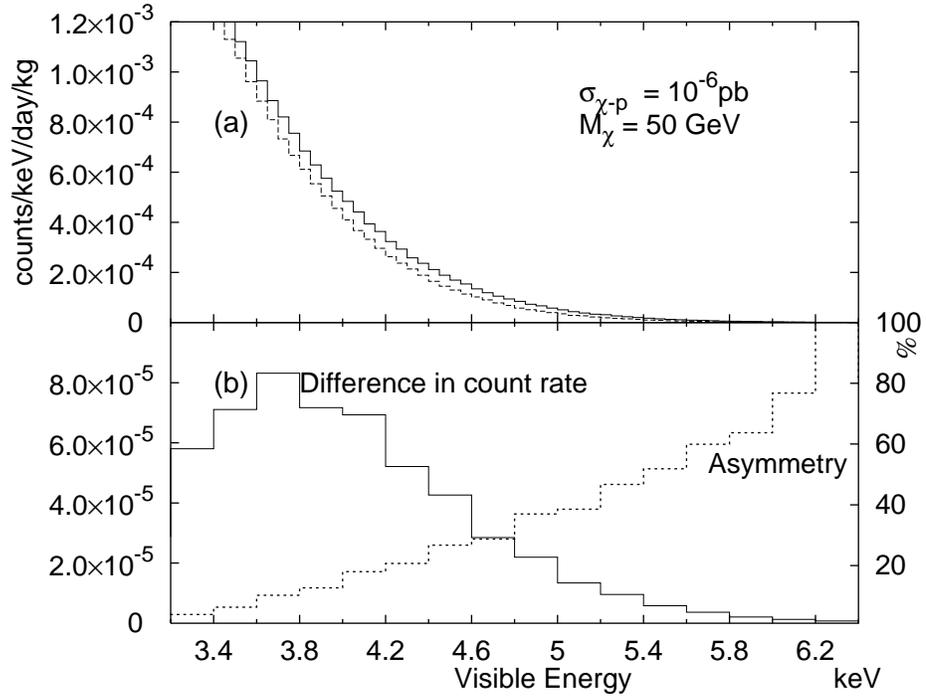}
\caption{(a) The expected energy spectra of the stilbene scintillator
 when $c'$ axis maintained to the direction of galactic center (solid) and
 galactic rotation (dashed). The parameters that we used in the calculation were 
 $\rho_0=0.3$ GeV/cm$^3$, $v_0=220$ km/s, $v_E=232$ km/s, 
 WIMP-proton spin independent cross section $\sigma_{\chi-p}=10^{-6}$ pb,
 and $M_{\chi}=50$ GeV. (b) Difference in count rates and the asymmetry
 (= difference/average). The experimental resolution is not taken into
 account in the plots. }
\label{fig:estimation}
\end{center}
\end{figure}

We calculate by Monte Carlo method in two cases,
$c'$ axis maintained to the direction of the galactic
rotation (the direction of the constellation Cygnus) and
to the galactic center (the direction of the constellation
Sagittarius).
$\theta$-dependence of $(dL/dx)$ is assumed to be trigonometric as 
observed in high energy region \cite{heckmann}.
 The expected spectra and the variation amplitude is shown in
Fig. \ref{fig:estimation}.
This implies that it is plausible to use stilbene crystals for WIMP search. 
 The variation of the spectra is much the same as that of annual modulation.
However, less systematic uncertainty of the measurements can be achieved 
because the direction of the crystal axis to the WIMP wind 
can be controlled.

Other organic crystals, such as anthracene and naphthalene
show stronger dependence of scintillation efficiency on direction 
to crystallographic axis \cite{naph}. These crystals should be more sensitive to
WIMP wind and the variation of the spectra should be larger. 
Especially, it was reported that the variation of the scintillation 
efficiency of naphthalene crystal is about 50 $\%$.
With these organic crystals, it would also be possible to obtain some
information on the motion of the galactic halo.

\section{Conclusion}
\label{conclusion}
We measured the scintillation efficiency of carbon recoils in a stilbene crystal.
The efficiency was observed to increase at low energies and depend
on the recoil direction with respect to the $c'$ axis.
The variation of the directional anisotropy is about 7 $\%$. 
Assuming this anisotropic response, we estimate the sensitivity to
the WIMPs and found that stilbene scintillator could detect
robuster WIMP signal arising from the earth's rotation around the
galactic center as compared with the annual modulation.

\section*{Acknowledgement}
\label{ack}
We would like to thank professor Masayuki Igashira and other members of
the Research Laboratory for Nuclear Reactors at Tokyo Institute of Technology for
allowing us to use the accelarator and their excellent operation.  

% \appendix

% \section{}
% \label{}

\end{document}